\documentclass[twocolumn,eqsecnum,aps,prb]{revtex4}

\usepackage{graphicx}%
\usepackage{dcolumn}
\usepackage{amsmath}

\makeatletter
\def\btt#1{\texttt{\@backslashchar#1}}%
\DeclareRobustCommand\bblash{\btt{\@backslashchar}}%
\makeatother

\begin{document}

\title{Novel features in the flux-flow resistivity of the heavy fermion superconductor PrOs$_{4}$Sb$_{12}$}

\author{M.~Kobayashi}

\author{H.~Sato}
\email{sato@phys.metro-u.ac.jp}

\author{H.~Sugawara}
\altaffiliation{Present Address: Faculty of Integrated Arts and Sciences, 
The University of Tokushima
Minamijosanjima-machi, Tokushima, 770-8502, Japan}

\author{H.~Fujiwara}
\author{Y.~Aoki}

\affiliation{Department of Physics, Tokyo Metropolitan University, Minami-Ohsawa, Hachioji, Tokyo 192-0397, Japan}

\date{\today}

\begin{abstract}
We have investigated the electrical resistivity of the heavy fermion superconductor PrOs$_{4}$Sb$_{12}$ in the mixed state. We found unusual double minima in the flux-flow resistivity as a function of magnetic field below the upper critical field for the first time, indicating double peaks in the pinning force density ($F_{\rm P}$). Estimated $F_{\rm P}$ at the peak exhibits apparent dependence on applied field direction; composed of two-fold and four-fold symmetries mimicking the reported angular dependence of thermal conductivity ($\kappa$). The result is discussed in correlation with the double step superconducting (SC) transition in the specific heat and the multiple SC-phases inferred from the angular dependence of $\kappa$.\\
\end{abstract}


\maketitle
PrOs$_{4}$Sb$_{12}$, a member of the filled skutterudite family with the tetrahedral ($T_{\rm h}$) point group symmetry,~\cite{Jeitschko} is the first Pr-based heavy fermion (HF) superconductor with a SC-transition temperature $T_{\rm C}=1.85$~K and an upper critical field $H_{\rm C2} =2.45$~T.~\cite{Bauer} Highly enhanced effective mass has been judged by the large specific heat jump $\Delta C/T_{\rm C}=500$~mJ/mol K$^{2}$,~\cite{Bauer,Maple} and more directly by the de Haas-van Alphen experiment.~\cite{Sugawara}

The Sb-NQR measurement has revealed that the temperature $T$ dependence of nuclear-spin-lattice-relaxation rate shows no coherence peak just below $T_{\rm C}$ but an exponential $T$ dependence,~\cite{Kotegawa} which is explainable as neither the conventional s-wave type nor any unconventional ones with the line-node gap. In the zero field specific heat~\cite{Maple,Aoki,Vollmer} and the dc magnetization,~\cite{Tayama} double step SC-transitions at $T_1$ and $T_2$ ($<T_1$) have been observed, which reminds us the two SC-phases in UPt$_3$.~\cite{Hasselbach}  Measson {\it et al.} have carefully determined two phase boundaries $T_{\rm 1}(H)$ and $T_{\rm 2}(H)$ by the ac specific heat measurement.~\cite{Measson}  Izawa {\it et al.} reported angular dependence of $\kappa$ with two different symmetries across a boundary in the $H-T$ phase diagram. They ascribed the symmetry change to the distinct SC-phases with the gap function $\Delta(k)$ having two point nodes in low fields and six point nodes in high fields.~\cite{Izawa} Huxley {\it et al.} found flux-line distortion consistent with the two-fold symmetry in $\kappa$ by the small angle neutron scattering experiment.~\cite{Huxley} Consistently, Chia {\it et al.} reported $T^2$-dependence in the penetration depth indicating the $T^2$-dependent superfluid density below 0.3$T_{\rm C}$,~\cite{Chia} from which point nodes on the Fermi surface have been inferred. In a recent muon spin relaxation experiment, Aoki {\it et al}. found spontaneous appearance of an internal magnetic field below $T_{\rm C}$.~\cite{Aoki_mSR} As intermetallic compounds, PrOs$_{4}$Sb$_{12}$ is the first superconductor with a time-reversal symmetry (TRS) breaking, since such SC-state with broken TRS has been confirmed only in Sr$_2$RuO$_4$ until now.~\cite{Mackenzie}  In discussing the origin of this attractive SC-state, the existence of field induced ordered phase (FIOP) close to the SC-state in the $H-T$ phase diagram is inferred to be important.~\cite{Aoki,Vollmer,Tayama,Ho,Rotundu} Recently, the FIOP has been ascribed to an antiferro-quadrupolar ordering based on neutron diffraction experiments.~\cite{Kohgi}

In order to have deeper insight into such an unconventional SC-state in PrOs$_{4}$Sb$_{12}$, we have investigated the flux-flow resistivity in the mixed state that is sensitive to the change in vortex state and energy dissipation due to quasiparticles trapped within voltex cores.~\cite{Tinkham,Larkin}  

Single crystals of PrOs$_{4}$Sb$_{12}$ were grown by Sb-self-flux method, using high-purity elements, 4N (99.99\% pure)-Pr, 4N-Os, and 6N-Sb.~\cite{Jeitschko,Sugawara} On the samples grown in the same manner and from the same raw materials, de Haas-van Alphen oscillations had been clearly detected,~\cite{Sugawara} which indirectly proves high quality of the samples. The electrical resistivity ($\rho$) under magnetic fields was measured by the conventional dc four-probe method. Error in determining angle between the current direction and the axis of field rotation is a few degrees. 

\begin{figure}[!tbp]
\begin{center}\leavevmode
\includegraphics[width=1\linewidth]{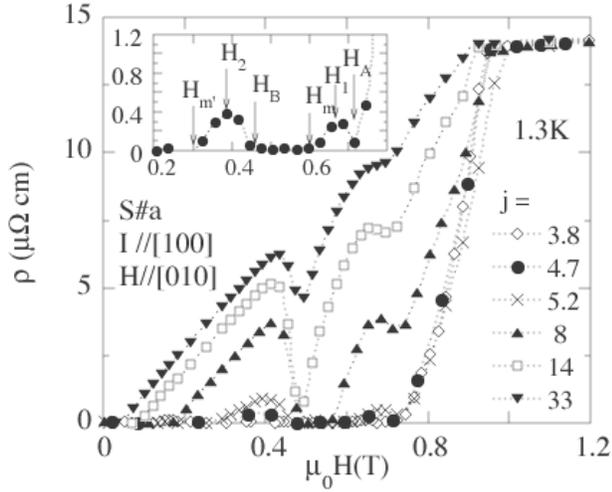}
\caption{Magnetic field dependence of electrical resistivity ($\rho$) in PrOs$_4$Sb$_{12}$ for selected values of current densities $j (10^{4}$ A/m$^{2}$). An expanded view for $j=4.7$ is shown in the inset where typical field names used in the text are defined.}
\label{fig1}
\end{center}
\end{figure}

Figure~\ref{fig1} shows the magnetic field $H$ dependence of $\rho$ at 1.3~K for selected values of current density $j$.
We plot only the field increasing curves in this paper for clarity, though slight hysteretic behaviors have been found. For $j$ smaller than $3.8\times 10^{4}$ A/m$^{2}$, the absence of flux flow manifests in zero resistivity up to $H_{\rm C2}$. Above $j=3.8\times 10^{4}$ A/m$^{2}$, $\rho$ starts to increase at a field $H_{\rm m}$ (e.g., $H_{\rm m}=0.58$~T for $j=3.8\times 10^{4}$ A/m$^{2}$, note the notation in the inset for $j=4.7\times 10^{4}$ A/m$^{2}$) where the Magnus force defined as $F_{\rm m}=|j\times H_{\rm m}|$ overcomes the pinning force. After reaching a peak at $H_1$, $\rho$ drops to zero at around $H_{\rm A}=0.70$~T, suggesting an enhanced pinning force at around $H_{\rm A}$. 

\begin{figure}[!tbp]
\begin{center}\leavevmode
\includegraphics[width=1\linewidth]{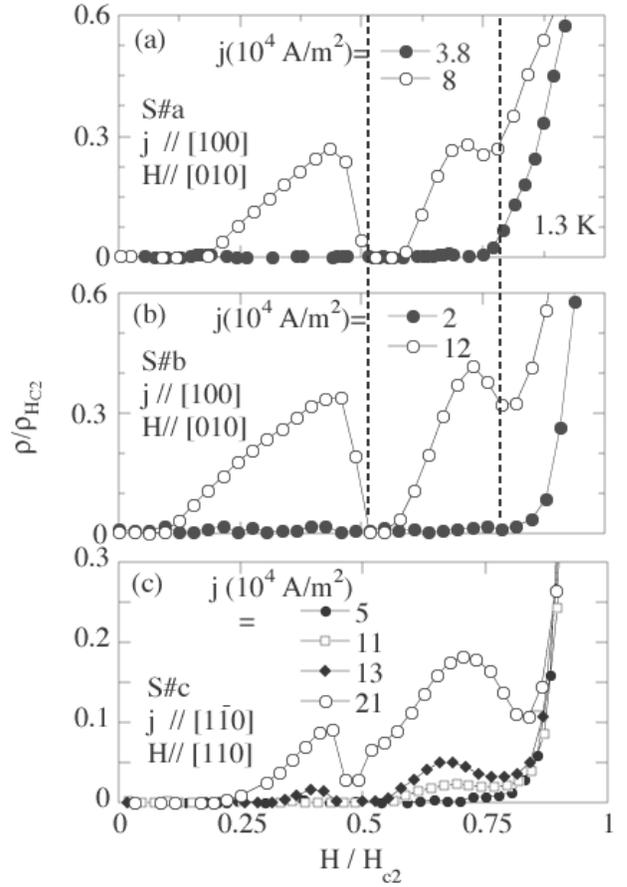}
\caption{Double minima structure in the flux flow resistivity. (a) and (b) show sample dependence for the same crystallographic configuration, and (c) for a different crystallographic orientation.}
\label{fig2}
\end{center}
\end{figure}

\begin{figure}[!tbp]
\begin{center}\leavevmode
\includegraphics[width=1\linewidth]{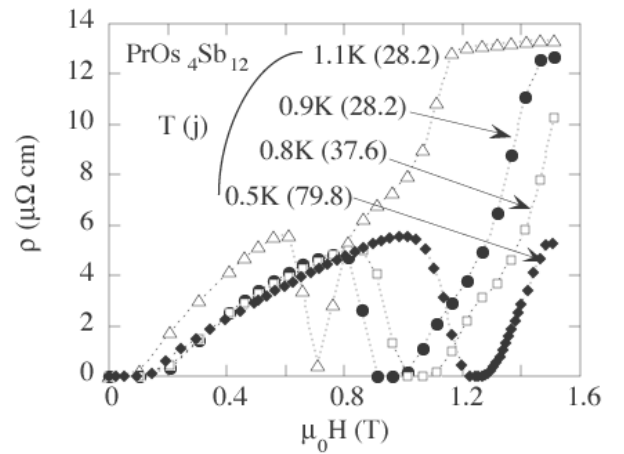}
\caption{Variation of the field dependence of flux-flow resistivity $\rho$ in PrOs$_4$Sb$_{12}$ at temperature $T$ and current density $j (10^4$ A/m$^2$).}
\label{fig3}
\end{center}
\end{figure}

For larger $j (=4.7\times 10^{4}$ A/m$^{2}$), the field $H_{\rm m'}=0.29$~T where $\rho$ starts to increase, is much smaller than that for $j=3.8\times 10^{4}$ A/m$^{2}$. After reaching a peak at $H_{2}=0.39$~T, $\rho$ drops to zero around $H_{\rm B}=0.48$~T. On further increasing $H$, $\rho$ shows a second peak at $H_1=0.69$~T, drops again around $H_{\rm A}=0.72$~T, and finally reaches the normal state resistivity at $H_{\rm C2}=0.96$~T. Double minima at around $H_{\rm A}$ and $H_{\rm B}$ in $\rho(H)$ suggest the successive enhancement of pinning force related with so-called peak effect in the magnetization. A minimum in $\rho(H)$ has been observed in many superconductors including ones containing rare earth ions.~\cite{Sato} In those experiments, however, only a single minimum has been reported just below $H_{\rm C2}$ until now, and such double minima structure as in Fig.~\ref{fig1} has not been reported before except our preliminary report to the best of our knowledge.~\cite{Kobayashi} In order to confirm the reproducibility, we have checked the sample dependence of $\rho(H)$ as shown in Fig.~\ref{fig2}, where both the resistivity and magnetic field are normalized at $H_{\rm C2}$ for the comparison.
The residual resistivity ratio RRR$=\rho_{\rm 300K}/\rho_{\rm 1.9K}$ is slightly different among the present samples; 27 for S\#a, 24 for S\#b and 34 for S\#c. The double peak structure was always observed with reproducible features such as the positions of $H_{\rm A}$ and $H_{\rm B}$ for the same crystallographic geometry; $j\|[100]$ and $H\|[010]$ as shown in Fig.~\ref{fig2}(a) and (b) except minor quantitative difference.  Also for different crystallographic geometries, we found basically the same double peak structure except some anisotropy as shown in Fig.~\ref{fig2}(c). The effect of $T$ on flux-flow resistivity is shown in Fig.~\ref{fig3} where both $H_{\rm A}$ and $H_{\rm B}$ increase with decreasing $T$.
Such an apparent $T$-dependence rules out the matching effect caused by $T$-independent alignment of pinning centers as an origin of this peak effect.  In the present case, so-called synchronization effect may be the most probable origin of the peak effect in PrOs$_{4}$Sb$_{12}$,~\cite{Larkin} though there remains an important question why the synchronization occurs at the two different magnetic field strengths.

\begin{figure}[!tbp]
\begin{center}\leavevmode
\includegraphics[width=0.85\linewidth]{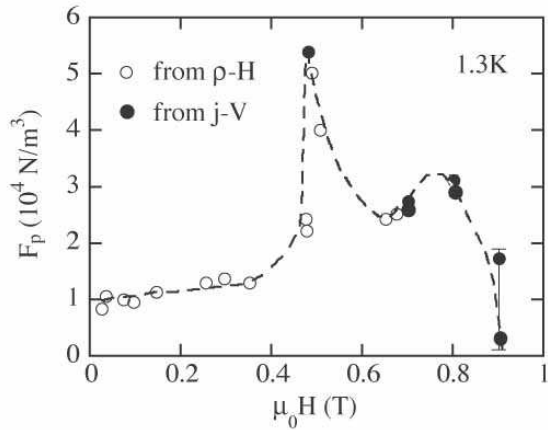}
\caption{Field dependence of the pinning force density $F_{\rm p}$ estimated from the field dependence of $\rho$ and $j-V$ characteristics for $j\|[100]$ and $H\|[010]$. }
\label{fig4}
\end{center}
\end{figure}

From the critical current density $j_{\rm C}$ where finite voltage starts to appear in the flux flow experiment, the pinning force density can be estimated as $F_{\rm p} =|j_{\rm C}\times H|$. The field dependence of the estimated $F_{\rm p}$ is shown in Fig.~\ref{fig4}.
$F_{\rm p}$ is $2.6\times 10^{4}$ N/m$^3$ at $H_{\rm A}=0.70$~T and $5.0\times 10^4$ N/m$^3$ at $H_{\rm B}=0.49$~T which are comparable with those of CeRu$_2$ ($\sim 3\times 10^5$ N/m$^3$) and UPt$_3$ ($\sim 1\times 10^4$ N/m$^3$).~\cite{Sato,Kambe}

\begin{figure}[!tbp]
\begin{center}\leavevmode
\includegraphics[width=1\linewidth]{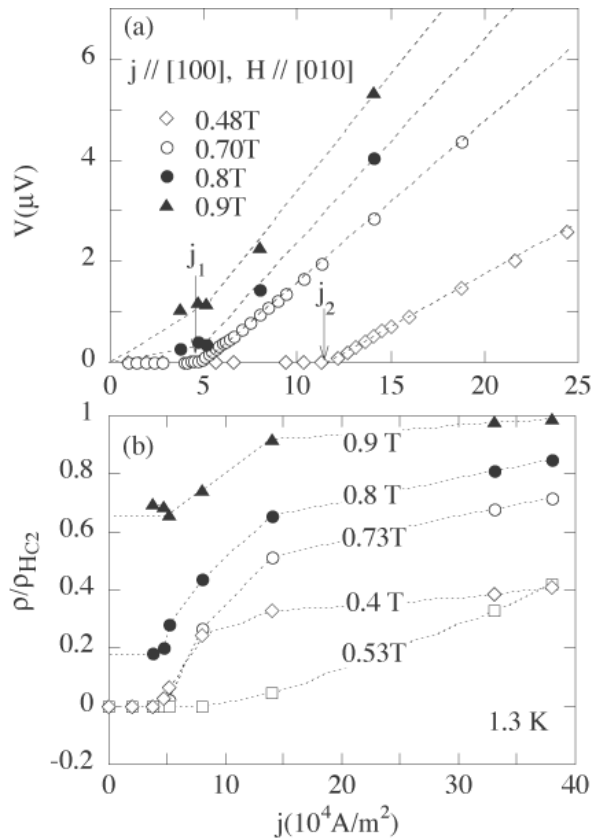}
\caption{(a) Typical $j-V$ characteristics under selected magnetic fields ($H_{\rm A}$ and $H_{\rm B}$) at 1.3~K. (b) the resistivity normalized at $H_{\rm C2}$ against current density transferred from Fig.~\ref{fig1}.}
\label{fig5}
\end{center}
\end{figure}

Taking into account the flux creep phenomenon, we have determined $F_{\rm p}$ more accurately by measuring the current density ($j$) - voltage ($V$) characteristics under constant fields. Typical $j-V$ curves for selected magnetic fields including 0.48~T ($H_{\rm B}$) and 0.70~T ($H_{\rm A}$) are plotted in Fig.~\ref{fig5}(a), where the $j-V$ curves under 0.8~T and 0.9~T taken from Fig.~\ref{fig1} are included.
Under 0.48~T, for example, the resistive voltage is zero up to $j_{\rm C}=1.1\times 10^5$ A/m$^2$, and starts to increase with increasing $j$. $F_{\rm p}$ estimated from this $j_{\rm C}$ is also plotted as solid circles in Fig.~\ref{fig4}. The figure suggests the qualitative feature of $F_{\rm p}(H)$ to be unchanged, although there exists minor contribution of the flux creep phenomenon to the $\rho(H)$ measurements. To clarify the low current density behavior, $\rho/\rho_{H_{\rm C2}}$ is plotted as a function of $j$ in Fig.~\ref{fig5}(b). Note that above $H_{\rm A}$, a finite and weakly $j$-dependent voltage appears already at small current densities depending on $H$, and starts to increase at around $j_1$. Namely, the resistive voltage consists of two components between $H_{\rm A}$ and $H_{\rm C2}$.

\begin{figure}[!tbp]
\begin{center}\leavevmode
\includegraphics[width=1\linewidth]{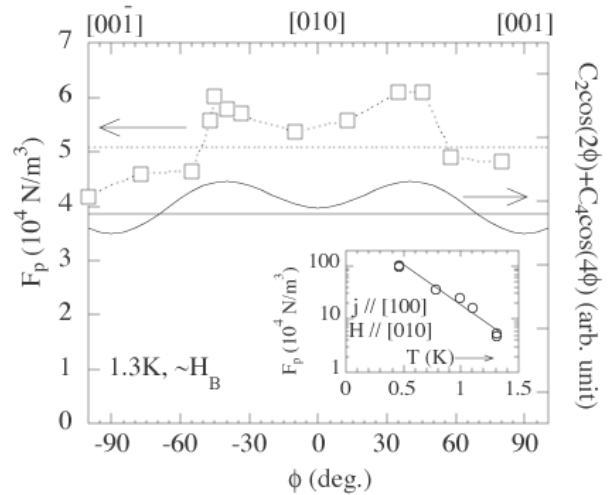}
\caption{Anglular dependence of the pinning force density $F_{\rm p}$ at $H_{\rm B}$. To exclude the effect of angular dependence of $H_{\rm B}$, $F_{\rm p}$ has been estimated at $H_{\rm B}$ for each field direction. The solid curve shows $\Delta k\sim 0.4\cos(2\phi)+0.6\cos(4\phi)$. Inset shows the $T$-dependence of $F_{\rm p}$ at $H_{\rm B}$.}
\label{fig6}
\end{center}
\end{figure}

The peak $F_{\rm p}$ estimated at $H_{\rm B}$ is found to depend on the field direction as shown in Fig.~\ref{fig6}.
Interestingly, the angular dependence mimics the anisotropy found in the thermal conductivity $\kappa(\phi)$ as a function of the angle $\phi$ of the applied magnetic field around the direction of thermal flow (particularly the curve for 0.78~T in Fig.~\ref{fig2} of Ref.~11). In the SC-state, $\kappa(\phi)$ was decomposed into three terms; isotropic, two-fold and four-fold symmetric ones. $\kappa$ shows four-fold symmetry above a critical field $H^*$, below which it changes into two-fold. The anisotropy was discussed in correlation with that in the SC-gap. The solid curve in Fig.~\ref{fig6} shows $\Delta F_{\rm p}(\phi)\sim 0.4\cos(2\phi)+0.6\cos(4\phi)$ which roughly reproduces the main features of $\Delta F_{\rm p}(\phi)$. $H_{\rm B}/H_{\rm C2}\approx0.5$ for the experimental condition in Fig.~\ref{fig6} is larger than the critical field $H^*/H_{\rm C2}\approx0.4$ in Ref.~11. It should be noted that the approximate four-fold-symmetry of $F_{\rm p}$ at $H_{\rm B}$ mimics that of $\kappa(\phi)$ in the high field phase in Ref.~11. The present finding of the anisotropy in $F_{\rm p}$ can be another proof of the anisotropic SC-state in PrOs$_{4}$Sb$_{12}$, and gives a new insight that the anisotropy must be related with the voltex core state as was clearly identified in the neutron scattering experiment.~\cite{Huxley} However, such a huge magnitude of oscillation in $F_{\rm P}$ ($\sim30\%$) compared to those in $\kappa(\phi)$~\cite{Izawa} and in a recent specific heat experiment~\cite{Custers} (a few percent) might not be simply ascribed only to the anisotropic quasiparticle excitations due to the anisotropic SC-gap, and theoretical studies are highly required to understand the mechanism leading to such a strong oscillation in $F_{\rm P}$. The inset in Fig.~\ref{fig6} shows the $T$-dependence of $F_{\rm p}$ just at $H_{\rm B}$. Within this temperature range, $F_{\rm p}$ becomes an order of magnitude enhanced with decreasing $T$, indicating the smallness of the potential barrier associated with the pinning.~\cite{Tinkham}

\begin{figure}[!tbp]
\begin{center}\leavevmode
\includegraphics[width=1\linewidth]{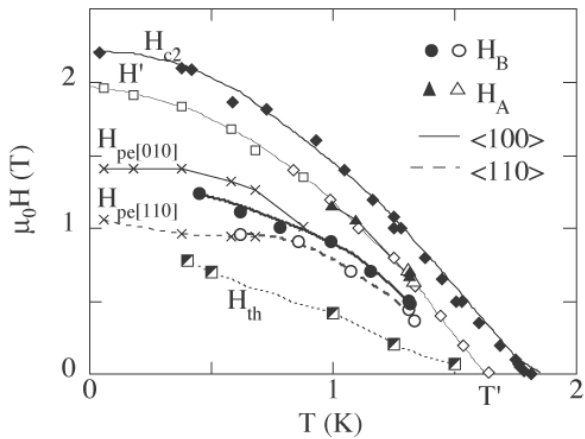}
\caption{$H-T$ phase diagram for the characteristic fields $H_{\rm A}$ and $H_{\rm B}$ estimated from the flux flow resistivity measurements.  $H_{\rm C2}$, $H$', $T$Õ, and $H_{\rm pe}$ are from the dc magnetization,~\cite{Tayama} and $H_{\rm th}$ by the thermal conductivity.~\cite{Izawa}}
\label{fig7}
\end{center}
\end{figure}

Finally, the fields $H_{\rm A}$ and $H_{\rm B}$ are plotted as a function of $T$ in the $H-T$ phase diagram (Fig.~\ref{fig7}), along with the related characteristic fields determined by the various experimental methods reported previously~\cite{Tayama,Izawa}; the transition points $H'$ and $T'$ observed in $dM(H)/dH$ and $dM(T)/dT$, and the onset field $H_{\rm pe}$ of the peak effect in the $M(H)$.~\cite{Tayama}
$H_{\rm th}$ is the phase boundary inferred from the $\kappa$ measurement.~\cite{Izawa} An important feature of this figure is the presence of $H_{\rm A}$ almost on $T_{2}(H)$; i.e., $F_{\rm p}$ has a peak close to the lower transition temperature. This fact can be a key feature to clarify the origin of the double peak structure in the specific heat which has not yet been settled, including whether it is intrinsic or not. On the other hand, the points $H_{\rm B}$ are roughly on the lines connecting $H_{\rm pe}$ in the magnetization measurement including the anisotropy, and clearly above $H_{\rm th}(T)$. In the dc-resistivity measurement, it is impossible to determine $F_{\rm p}$ at lower temperatures, since the stronger pinning force requires the higher current density.  It should be necessary to utilize surface impedance experiment to clarify the lower temperature voltex state of this exotic SC-material.

In summary, we have found the double minima structure in the flux flow resistivity.  The higher field one appears close to the field of the lower temperature SC-transition in the specific heat, while the lower field one appears at a field above the boundary of the two SC-phases suggested by the $\kappa$-measurements.  The two-fold symmetry component reported in the angular dependence of $\kappa$ and neutron scattering experiment has been also found in the angular dependence of the pinning force density superposed upon the four-fold symmetry one.

\begin{acknowledgments}
We thank Prof. H. Harima, Prof. Y. Matsuda, Dr. K. Izawa, Prof. T. Tamegai, Prof. M. Ichioka, Prof. K. Machida, Prof. M. Kohgi, Prof. O. Sakai and Dr. Shiina for helpful discussion.  This work was supported by a Grant-in-Aid for Scientific Research Priority Area "Skutterudite" (No.15072206) of MEXT, Japan.
\end{acknowledgments}

\end{document}